\documentclass[aps,pra,reprint,twocolumn,superscriptaddress]{revtex4-1}
\usepackage{graphicx}

\begin{document}

\title{Quantum and classical telecommunication channel multiplexing}

\author{Andreas Lenhard}
\affiliation{Quantenoptik, Fachrichtung 7.2, Universit\"at des Saarlandes, Campus E2.6, 66123 Saarbr\"ucken, Germany}

\author{Jos\'e Brito}
\author{Stephan Kucera}
\affiliation{Quantenphotonik, Fachrichtung 7.2, Universit\"at des Saarlandes, Campus E2.6, 66123 Saarbr\"ucken, Germany}

\author{Matthias Bock}
\affiliation{Quantenoptik, Fachrichtung 7.2, Universit\"at des Saarlandes, Campus E2.6, 66123 Saarbr\"ucken, Germany}

\author{J\"urgen Eschner}
\affiliation{Quantenphotonik, Fachrichtung 7.2, Universit\"at des Saarlandes, Campus E2.6, 66123 Saarbr\"ucken, Germany}

\author{Christoph Becher}
\email[]{christoph.becher@physik.uni-saarland.de}
\affiliation{Quantenoptik, Fachrichtung 7.2, Universit\"at des Saarlandes, Campus E2.6, 66123 Saarbr\"ucken, Germany}

\begin{abstract}
We demonstrate the multiplexing of a classical coherent and a quantum state of light in a single telecommunciation fiber. For this purpose we make use of spontaneous parametric down conversion and quantum frequency conversion to generate photon pairs at 854~nm and the telecom O-band. The herald photon triggers a telecom C-band laser pulse. The telecom single photon and the laser pulse are combined and coupled to a standard telecom fiber. Low background time correlation of the classical and quantum signal behind the fiber shows successful telecommunication channel multiplexing.
\end{abstract}


\maketitle

\section*{Introduction}\label{intro}
Today's internet traffic that is distributed over optical fiber channels mainly concentrates on the telecom C- (1530 - 1565~nm) and L-band (1565 - 1625~nm) because in-line active amplifiers based on erbium-doped fibers offer a feasible way to build repeaters for the signals. Hence there exist a number of telecom frequency bands which are not used, being available for quantum communication.

In state of the art realizations, quantum key distribution (QKD) is possible over 336~km \cite{Shi14} and entanglement distribution was demonstrated with lengths up to 300~km \cite{Ina13}. QKD with such high loss channels became possible only with the development of low-noise detectors \cite{Tak07}. Based on this progress, QKD over existing fiber networks in urban areas has been demonstrated \cite{Pee09,Sas11,Rub13} as well as entanglement distribution \cite{Tit98} and teleportation \cite{Mar03}.

QKD protocols, e.g. BB84 \cite{BB84}, usually involve an exchange of data via a quantum and a classical channel simultaneously to establish the final secret key. Realizing both channels in a single fiber could reduce the device complexity. In first experimental demonstrations, weak coherent pulses at the single photon power level in the telecom O-band (1260 - 1360~nm) were multiplexed with a common data channel in the C-band and sent via several kilometers of installed fiber \cite{Tow97} with an acceptable error rate of the quantum channel. The dominant noise source in that scheme is anti-Stokes Raman scattering of the strong classical pulses into the quantum channel \cite{Cha09}. Furthermore, it is also possible to transmit the O-band photons through a fiber link equipped with active erbium amplifiers. The spontaneous emission of the amplifiers at O-band wavelengths is quite low and the amount of additional noise photons in the quantum channel can be neglected after narrow spectral filtering \cite{Hal09}. However, each amplifier attenuates the O-band photons (e.g. 11~dB loss in \cite{Hal09}). Nevertheless, in such a configuration the distribution of entangled states was demonstrated \cite{Hal09}.

The telecom C-band is divided into a grid of equally spaced channels for dense wavelength multiplexing (12.5~GHz - 100~GHz spacing in ITU-grid \cite{Itu12}). Each of these channels can be used independently for classical communication and filters are available to separate and demultiplex individual channels. However, the suppression of such common filters is not high enough (20-40~dB) to protect a quantum channel from neighboring classical states of light. If such a fiber link includes an amplifier, a quantum channel in the C-band will also be classically amplified which destroys its quantum state (no cloning theorem). From these considerations, transmitting the quantum channel in the O-band (taking losses into account) and classical data in the C-band seems the best solution.

To establish a quantum network, quantum nodes like single trapped atoms or ions, emitting in the visible or near infrared spectral region, have to be interconnected via photons. To this end, bridging the gap between low loss telecom wavelengths for long-range communication and the atomic wavelengths quantum frequency conversion (QFC) is necessary \cite{Ou08,Zas12,Blu13}. In the present experiment we produce frequency-degenerate photon pairs, generated by spontaneous parametric down conversion (SPDC), resonant with an atomic transition of $^{40}$Ca$^+$ at 854~nm. One of these photons is frequency converted to the telecom O-band. Its partner photon triggers a telecom C-band laser pulse which is overlapped with the converted photon and transmitted through a long single mode fiber. We measure non-classical correlations between the laser pulse and the converted photon at the fiber output, which arise from the original time-correlation of the SPDC pair, maintained along frequency conversion, multiplexing and fiber transmission.

\section*{Fiber Characterization}
\begin{figure}[tb]
	\centering
		\includegraphics[width=0.45\textwidth]{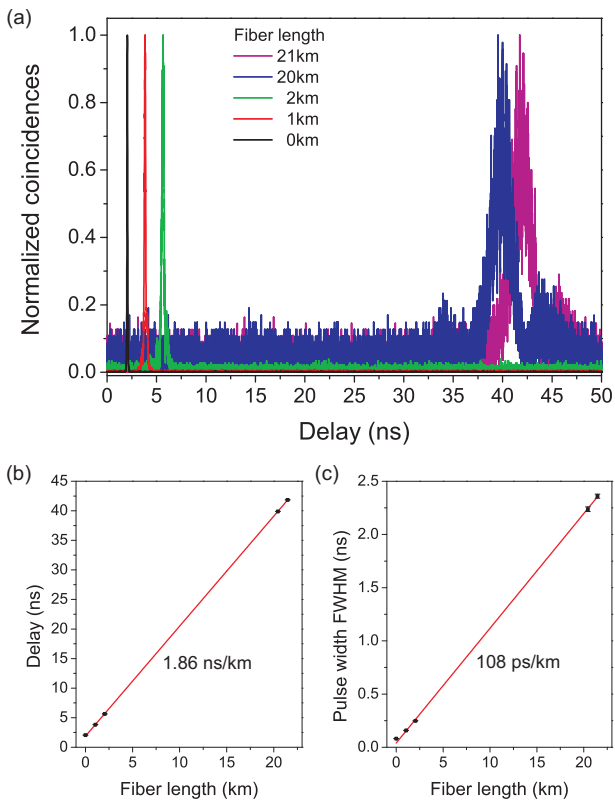}
	\caption{Dispersion effects in the telecom fiber. a) shows correlation measurements between O-band and C-band photon pairs sent through different fibers. b) shows the shift of the coincidence peak and c) the broadening of this peak for the different fibers.}
	\label{fig:FiberDispersion}
\end{figure}

To examine the different properties of fibers in the O- and C-band in a preliminary experiment, we use correlated photon pairs in each of these bands. For this purpose, we pump a lithium niobate waveguide \cite{Zas12} at a wavelength of 708~nm to generate SPDC photons at 1535~nm and 1313~nm with bandwidths of approximately 1~THz, respectively (details about SPDC in this particular device can be found in \cite{Boc15}). 
Both photons are sent through long single mode fibers (SMF-28e) before separating them with a wavelength division multiplexer (WDM). Each output of the WDM is connected to a single photon detector. In the post processing of the data we correlate the arrival times of the photons to measure the coincidences between O-band and C-band photons. The fiber has a different refractive index in the O- and C-band and thus the delay between the photons will depend on the length of the fiber. Hence the position of the coincidence peak will move. The result is shown in Fig.~\ref{fig:FiberDispersion}.
Furthermore, even the individual spectral components of each photon will experience different phase velocities leading to pulse broadening. The dispersion minimum of SMF-28 fibers is in the telecom O-band ($1310<\lambda_0<1324$~nm, \cite{Cor07}). Hence we expect negligible pulse broadening there but the temporal shape of the C-band photon will significantly broaden due to dispersion. The results of the correlation measurement show that the delay between the O-band and C-band photons increases by $1.861\pm0.004$~ns/km. This value was measured at the peak positions and thus holds for every pulse-pair with 1535~nm and 1313~nm central wavelength. For the pulse broadening of the C-band photons we find a value of $108\pm0.8$~ps/km which depends on the particular spectral shape.

In practical realizations such effects can hamper the applicability. In internet communication the link between two nodes can be reconfigured on demand to reduce data traffic in certain channels and optimize speed and workload. Hence the optical distance between two nodes might change. For the idea of multiplexing quantum and classical channels this means the delay between these data packets might also change. 
There exist techniques to track the signal via correlation measurements \cite{Ho09} which then need to be implemented, increasing complexity.

\section*{Quantum and Classical Channel in a Single Fiber}
\begin{figure*}[tb]
	\centering
		\includegraphics[width=0.95\textwidth]{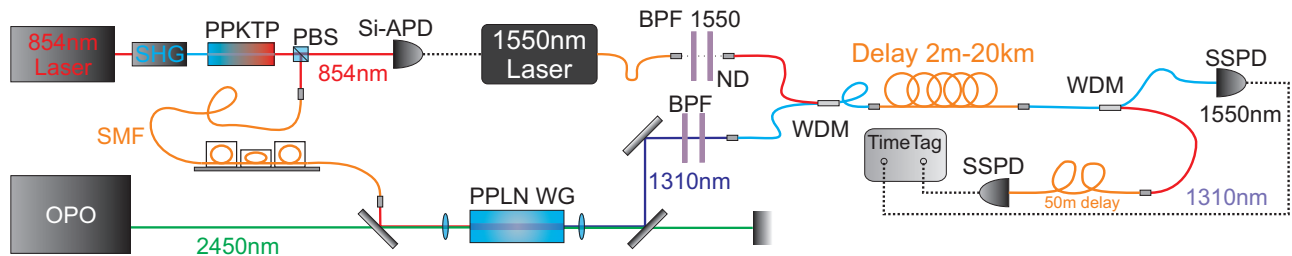}
	\caption{Experimental setup for correlation measurements between converted photons and an optical clock signal. For abbreviations see text.}
	\label{fig:SPDCConversionOptickClock-Setup}
\end{figure*}

To demonstrate the multiplexing of two optical signals in a single fiber (SMF-28e), we implemented the experiment described in the following and illustrated in Fig.~\ref{fig:SPDCConversionOptickClock-Setup}.

We use spontaneous parametric down conversion (SPDC) to generate photon pairs at 854~nm. This source has been described earlier in more detail \cite{Pir09,Pir10} where it has been used to herald the absorption of a single-photon by a single trapped ion. It consists of a diode laser system at 854~nm which is actively stabilized to an atomic transition in calcium ($^{40}$Ca$^+$, $3^2$D$_{5/2}$$\leftrightarrow$$4^2$P$_{3/2}$). The second harmonic (SHG) of this laser field serves as pump field for the SPDC process. This allows to generate frequency-degenerate photons resonant with the transition in Ca$^+$. The down conversion is realized via type-II phase matching conditions in a 2~cm long periodically poled KTP (PPKTP) crystal. The photons are thus polarization entangled. However, for the experiments reported here this kind of entanglement is not used and we employ a polarizing beam splitter (PBS) to effectively separate the signal and idler photons. One photon serves as a herald, which is detected by a silicon avalanche photon detector (Si-APD, Perkin Elmer SPCM-AQR-14, 30~\% detection efficiency at 854~nm), while its partner photon is sent to the frequency converter setup.
For frequency conversion we make use of difference frequency generation in a periodically poled lithium niobate ridge waveguide (similar to the device described in \cite{Zas12}). The conversion from $\lambda_s=854$~nm to the telecommunications O-band around $\lambda_i=1310$~nm is stimulated by a strong coherent pump field at a wavelength of $\lambda_p=2453$~nm ($1/\lambda_s-1/\lambda_p=1/\lambda_i$). This pump field is generated by a continuous wave optical parametric oscillator (OPO). From calibration measurements we determine the over-all conversion efficiency to approximately 8~\% (including coupling of signal photons to the waveguide and telecom photons to the output fiber). Finally, we use superconducting single photon detectors (SSPD, Single Quantum EOS X10) for counting the telecom photons (detection efficiency of 25~\% at 1310~nm).
Although the efficiency of frequency conversion is only 8~\% here, the overall transmission efficiency (including conversion and fiber transmission losses) benefits from the small fiber losses in the O-band (measured as 0.25~dB/km at 1310~nm, compared to approx. 2~dB/km at 854~nm) and equals the transmission efficiency of the unconverted photons at a fiber length of 6.5~km. After 20~km of fiber the advantage of the conversion scheme amounts to 24~dB.
These numbers illustrate that low-loss transmission of quantum states through fibers benefits from QFC.

\begin{figure*}[htb]
	\centering
		\includegraphics[width=0.95\textwidth]{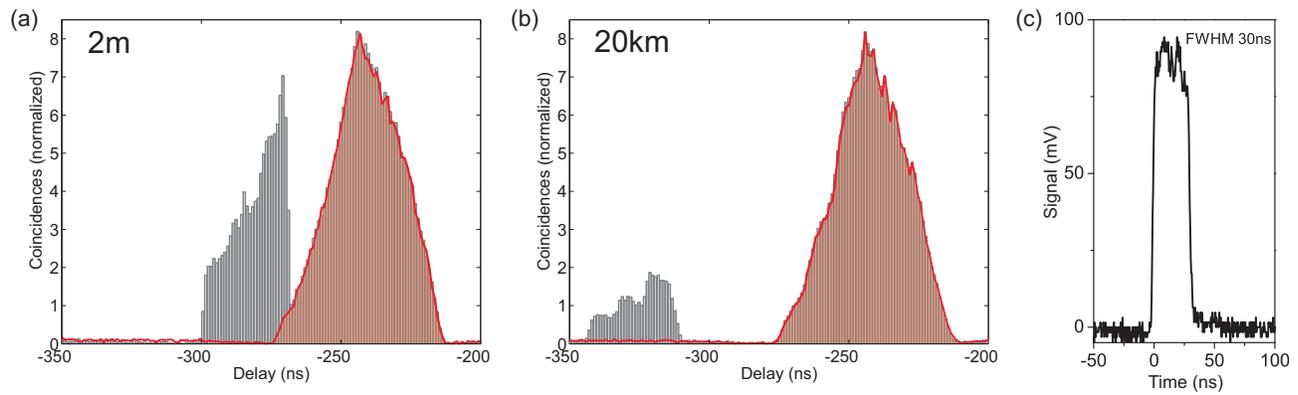}
	\caption{Converted single photons at telecom O-band and laser herald pulses in the telecom C-band multiplexed in a single fiber. a) shows the results for 2~m of delay, as reference, and b) shows the long-range measurement over 20~km. The red solid lines show measurements of the background. The red shaded areas are a guide for the eye and indicate the cross-talk coincidences of the laser pulses. c) shows a measurement of the laser pulse intensity with a standard photo diode and an oscilloscope (2~m fiber delay).}
	\label{fig:SinglePhoton-LaserClock}
\end{figure*}

To multiplex classical and quantum states of light in a single fiber, we connect the heralding detector (Si-APD) to a diode laser (Thorlabs LPSC-1550-DC) emitting at 1550~nm in the telecom C-band. In detail the diode was biased with a DC current of 11~mA and the TTL-pulses were added via a bias-T and a 50~Ohm load resistor. This is enough to drive the diode above threshold (36~mA). The shape of the generated optical pulse (Fig.~\ref{fig:SinglePhoton-LaserClock}c) follows the electrical current. The laser pulse is transmitted through 1550~nm bandpass-filters (BPF, 20~nm FWHM) to suppress noise at other telecom wavelengths, generated by the laser. A set of neutral density filters (ND) is used here to reduce the power. Then the pulses are sent to the lab where the converter is hosted. There, the laser pulses are combined with the converted photons via a WDM as shown in Fig.~\ref{fig:SPDCConversionOptickClock-Setup}.

After combination, both optical fields travel in the same delay fiber which can be set to various lengths between 2~m and 20~km. After this fiber link both fields are again separated by a single WDM. These WDMs (Thorlabs WD202B-FC) have a minimal isolation of 16~dB between the two bands. Both output channels are connected to the SSPDs. The O-band channel has an additional delay of 50~m of fiber to reduce cross-talk effects in detection. In a preparatory measurement the rate of herald events was measured directly at the Si-APD. Then the laser pulses were attenuated by six orders of magnitude which, for the heralding channel, resulted in a comparable count rate on the SSPD as on the Si-APD before. We collect the detection events on both SSPD in a list of time-stamps using a fast counting electronics (Roithner Laser TTM8000) and correlate both later. For comparison a background measurement was performed by blocking the input of 854~nm photons in the converter and repeating the measurements with the same settings (red curves in Fig.~\ref{fig:SinglePhoton-LaserClock}). When the laser pulses arrive at the WDM the major part is directed to the C-band output port. However, a small fraction of the pulse (given by the WDM isolation) will also be directed to the O-band port. As the coherent pulse consists of many photons this leakage results in a coincidence detection on both detectors. The shape of this coincidence peak is given by the auto-correlation of the laser pulse and the position of the peak is given by the difference in optical delay between the WDM and the individual detectors. The background measurement makes this effect visible as the true single photon signal is blocked. Furthermore, we expect a very low constant noise floor by accidental coincidences between telecom O-band noise photons generated in the converter and C-band spontaneous emission of the laser, as well as detector dark counts in both channels. The laser pulses have a rectangular shape of 30~ns width (see Fig.~\ref{fig:SinglePhoton-LaserClock}c). Hence we expect a triangular auto-correlation with a width of 60~ns at the base. This corresponds well with the experimental width of 62~ns.
The results for simultaneous transmission of classical and quantum signals are shown in Fig.~\ref{fig:SinglePhoton-LaserClock}(a,b) for the shortest and longest available fiber lengths, respectively.
In both examples we can identify two coincidence peaks. We inserted an additional 50~m delay fiber between the two SSPD channels (measured to introduce 244~ns delay). Hence we can identify the peak around -250~ns to stem from the fraction of the C-band laser pulse leaking into the O-band detection channel. This is proved by the background coincidence rate (red lines in Fig.~\ref{fig:SinglePhoton-LaserClock}) which shows that this peak is completely independent of the frequency converted photons. In contrast, the neighboring peak (grey shaded area) stems from the correlation between the laser pulses and the converted single photons. The center of mass of this peak shifts by 43~ns which is comparable to the shift expected from the earlier results (37.2~ns for 20~km, see Fig.~\ref{fig:FiberDispersion}). For the correlation signal of interest we expect a convolution between the short photon pulse (coherence time 5~ps) and the laser pulse (30~ns). The measured width at the base of the 20~km delayed pulse is 34~ns, corresponding well with the estimated value (30~ns + 20~km$\cdot$108~ps/km = 32~ns). We can attribute the modulation of the pulse height to saturation and dead-time effects in the detectors. The coherent laser pulse consists of more than one photon on average which means that possibly more than one detection event can be triggered by a single pulse on the SSPD. In particular, the dead time of the SSPD is around 10~ns which is smaller than the pulse width, giving rise to the three-fold peak structure of the 30~ns long coincidence peak.

An effective separation of the signal of interest and the background is possible by gated detection, where the position and width of the gate window are adapted to position and shape of the signal coincidence peak.

\section*{Conclusion}
To summarize the results, we have shown a basic experiment demonstrating a way to multiplex classical and quantum channels in a single fiber. We demonstrated the correlation of the SPDC photons by transmitting the timing information by an optical clock pulse in the telcom C-band. Moreover, we used quantum frequency conversion to the telecom O-band to tackle the high transmission losses of quantum-memory compatible photons. Both QFC and mapping of the detection event to a laser pulse, preserve the time correlation of the original 854~nm partner photons. This is an important proof-of-principle demonstration for distribution of quantum information using existing fiber network infrastructure.

Regarding a practical application, we should use gated detection and choose the gate width and position to cut out the coincidence peak with the true signal only, thereby discarding the detection channel cross-talk effects we observed. For fibers with several kilometers of length the delay is large enough to make this feasible.

We also performed this measurement with non-attenuated laser pulses and additional filters before the single photon detector. In this case we were able to detect the laser pulses with a standard photodiode (Thorlabs DET10C), thus proving that single photon sensitivity is not necessary in the clock channel. Unfortunately, the background in the single-photon channel drastically increases. The experimental setup could be optimized in both cases by including narrowband filters (e.g. fiber bragg gratings) and WDMs with higher isolation. A further improvement is to use much shorter laser pulses that decrease timing jitter as well as reduce dead time effects in detection.

\section*{Acknowledgments}
The work was funded by the German Federal Ministry of Science and Education (BMBF) within the projects "Q.com-Q" (contract No.\ 16KIS0127). J.~Brito acknowledges support by CONICYT.

%

\end{document}